\documentclass[sigconf,authorversion,nonacm]{acmart}

\usepackage{tablefootnote}

\AtBeginDocument{%
  }

\setcopyright{acmlicensed}
\copyrightyear{2018}
\acmYear{2018}
\acmDOI{XXXXXXX.XXXXXXX}

\acmConference[Conference acronym 'XX]{Make sure to enter the correct
  conference title from your rights confirmation emai}{June 03--05,
  2018}{Woodstock, NY}
\acmISBN{978-1-4503-XXXX-X/18/06}

\begin{document}

%\title{On the intricacies of building, evaluating and deploying ranking models for Retrieval-Augmented Generation}
%\title{Improving RAG: building ranking models from encoder and decoder-based language models}
%\title{Improving text retrieval for RAG with ranking models}
\title{Enhancing Q\&A Text Retrieval with Ranking Models: Benchmarking, fine-tuning and deploying Rerankers for RAG}

\author{Gabriel de Souza P. Moreira}
\affiliation{%
  \institution{NVIDIA}
  \city{S\~ao Paulo}
  \country{Brazil}}
\email{gmoreira@nvidia.com}

\author{Ronay Ak}
\affiliation{%
  \institution{NVIDIA}
  \city{Sarasota}
  %\state{Beijing Shi}
  \country{USA}}
\email{ronaya@nvidia.com}

\author{Benedikt Schifferer}
%\email{webmaster@marysville-ohio.com}
\affiliation{%
  \institution{NVIDIA}
  \city{Berlin}
  \country{Germany}
}
\email{bschifferer@nvidia.com}

\author{Mengyao Xu}
\affiliation{%
 \institution{NVIDIA}
 \city{Santa Clara}
 %\state{Arunachal Pradesh}
 \country{USA}}
\email{mengyaox@nvidia.com}  

\author{Radek Osmulski}
\affiliation{%
  \institution{NVIDIA}
  \city{Brisbane}
  \country{Australia}
}
\email{rosmulski@nvidia.com} 

\author{Even Oldridge}
\affiliation{%
  \institution{NVIDIA}
  \city{Vancouver}
  %\state{Texas}
  \country{Canada}}
\email{eoldridge@nvidia.com}

\renewcommand{\shortauthors}{Moreira et al.}

\begin{abstract}

Ranking models play a crucial role in enhancing overall accuracy of text retrieval systems. These multi-stage systems typically utilize either dense embedding models or sparse lexical indices to retrieve relevant passages based on a given query, followed by ranking models that refine the ordering of the candidate passages by its relevance to the query. 

This paper benchmarks various publicly available ranking models and examines their impact on ranking accuracy. We focus on text retrieval for question-answering tasks, a common use case for Retrieval-Augmented Generation systems. Our evaluation benchmarks include models some of which are commercially viable for industrial applications.

We introduce a state-of-the-art ranking model, \textit{NV-RerankQA-Mistral-4B-v3}, which achieves a significant accuracy increase of ~14\% compared to pipelines with other rerankers. We also provide an ablation study comparing the fine-tuning of ranking models with different sizes, losses and self-attention mechanisms.

Finally, we discuss challenges of text retrieval pipelines with ranking models in real-world industry applications, in particular the trade-offs among model size, ranking accuracy and system requirements like indexing and serving latency / throughput.

\end{abstract}


%%\ccsdesc[500]{Information systems~Retrieval models and ranking}


\maketitle

\section{Introduction}
Text retrieval is a core component for many information retrieval applications like search, Question-Answering (Q\&A) and recommender systems. 
More recently, text retrieval has been leveraged by Retrieval-Augmented Generation (RAG)\cite{lewis2020retrieval,ram2023context} systems, empowering Large Language Models (LLM) with external and up-to-date context.

Text embedding models represent variable-length text as a fixed dimension vector that can be used for downstream tasks. They are key for effective text retrieval, as they can semantically match pieces of textual content that can be symmetric (e.g. similar sentences or documents) or asymmetric (question and passages that might containing its answer).

Embedding models are based on the Transformer architecture. Some examples of seminal works are Sentence-BERT \cite{reimers2019sentence}, DPR \cite{karpukhin2020dense}, E5 \cite{wang2022text} and E5-Mistral \cite{wang2023improving}. They are typically trained with Constrastive Learning \cite{chen2020simple, karpukhin2020dense} as a \textit{bi-encoder} or \textit{late combination model} \cite{zamani2018neural}, i.e. query and passage are embedded separately and the model is optimized to maximize the similarity between query and relevant (positive) passages and minimize the similarity between query and non-relevant (negative) passages.

Retrieval systems that leverage text embedding models typically split the corpus into small chunks or passages (e.g. sentences or paragraphs), embed those passages and index corresponding embeddings into a vector database. This setup allows efficiently retrieving relevant passages from the embedded query by using Maximum Inner Product Search (MIPS) \cite{lewis2020retrieval} or another Approximate Nearest-Neighbor (ANN) algorithm.

The MTEB \cite{muennighoff2022mteb} is a popular benchmark of text embedding models for different tasks like retrieval, classification, clustering, semantic textual similarity, among others. We can notice from MTEB leaderboard\footnote{\url{https://huggingface.co/spaces/mteb/leaderboard}} that in general the larger the embedding model in terms of parameters the higher the accuracy it can achieve. 
However, that brings engineering challenges for companies in deploying such systems, as large embedding models can be prohibitively costly or slow to index very large textual corpus / knowledge bases.

\begin{figure}[ht]
    \centering
    \includegraphics[width=0.7\linewidth]{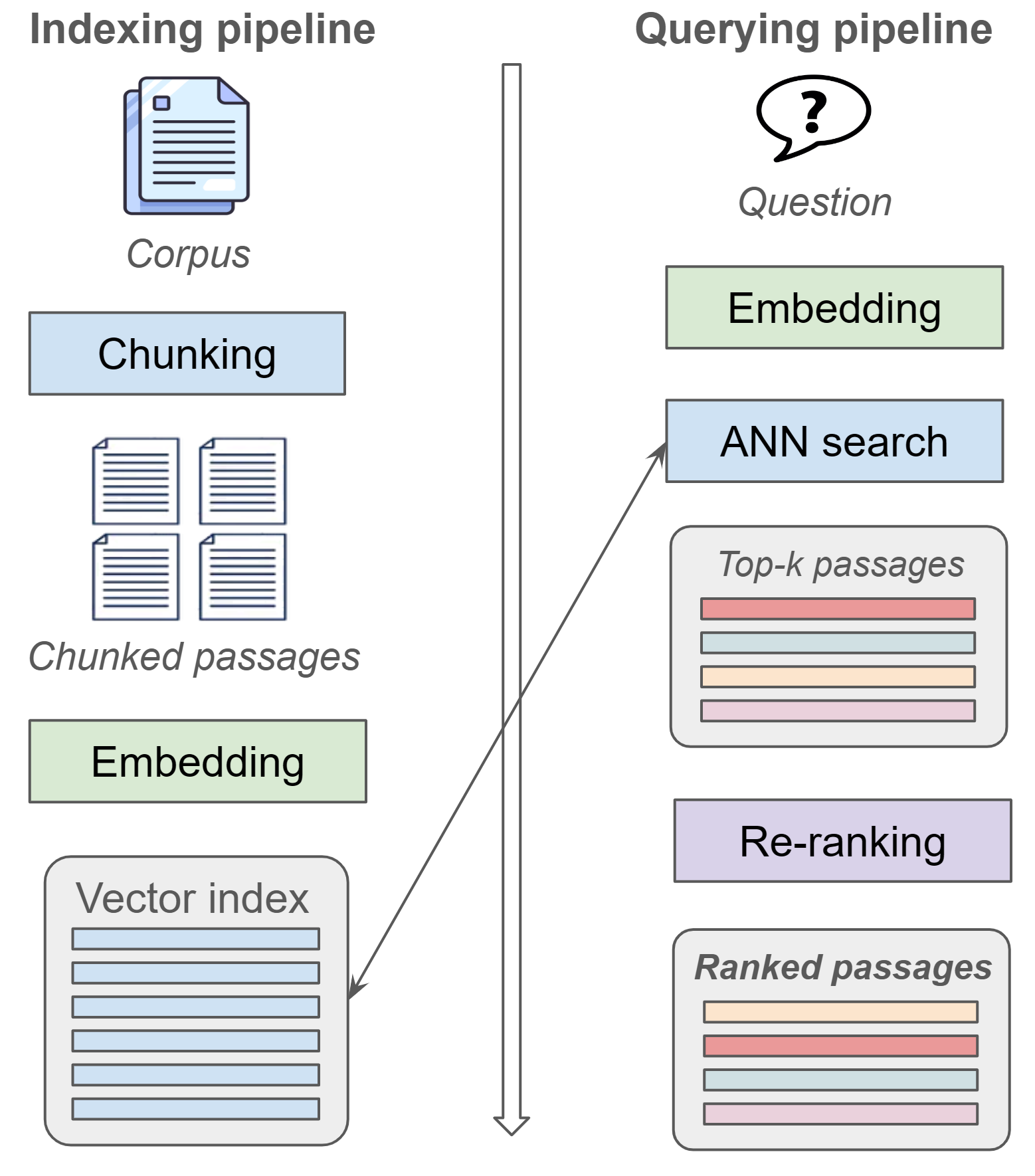}
    \caption{Illustration of the typical indexing and querying pipelines for multi-stage text retrieval}
    \label{fig:pipelines}
\end{figure}

For that reason, multi-stage text retrieval pipelines have been proposed to increase indexing and serving throughput, as well as improving the retrieval accuracy. In those pipelines, a sparse and/or dense embedding model are first used to retrieve top-k candidate passages, followed by a ranking model that refines the final ranking of those passages, as illustrated in Figure~\ref{fig:pipelines}.

Ranking models are typically Transformer models that operate as a \textit{cross-encoder} or \textit{early combination model} \cite{zamani2018neural} that takes as input both the query and passage pair concatenated and uses the self-attention mechanism to interact more deeply with the query and passage pair, and model their semantic relationship. Ranking models are used to provide relevance predictions only for the top-k candidates retrieved by the retrieval model. They can increase retrieval accuracy and make it possible using smaller embedding model, considerably reducing the indexing time and cost.

In this paper, we present a benchmark of publicly available ranking models for text retrieval and discuss how they affect the ranking accuracy compared to the original ordering provided by different embedding models. To ensure the usefulness of this benchmark for companies, in our experimentation we include certain embedding and ranking models that can be used commercially, i.e. that have a proper license and were not trained on research-only public data such as the popular MS-MARCO \cite{bajaj2016ms} dataset.

The main contributions of this paper are:

\begin{itemize}   
    \item We provide a comprehensive accuracy evaluation of publicly available ranking models with different commercially usable embedding models for Q\&A Text Retrieval;    
    \item We introduce a state-of-the-art ranking model: \textit{NV-RerankQA-Mistral-4B-v3}, and describe its architecture, pruning from Mistral 7B and fine-tuning techniques for leading ranking accuracy;
    \item We present an ablation study fine-tuning ranking models on top of different base model sizes. Then for the Mistral base model, we experiment with both point-wise and pair-wise losses and different attention/pooling mechanisms and discuss how these combinations affect the accuracy;
    \item Finally, we discuss trade-off aspects related to deploying text retrieval pipelines with or without a ranking model, like inference latency and embeddings indexing throughput.
\end{itemize}

\section{Background and Related work}

In earlier days of information retrieval, a common approach to refine search results based on sparse retrieval models (e.g. BM25) was to use feature-based learning-to-rank models with point-wise, pair-wise or list-wise losses \cite{liu2009learning}.

Neural Ranking Models (NRM) \cite{zamani2018neural}, typically feed-forward networks, were proposed to capture interactions between query and document. DeepMatch \cite{lu2013deep} modeled each text as a sequence of terms to generate a matching score. DRMM \cite{guo2016deep} represents the input texts as histogram-based features. Duet \cite{mitra2017learning} proposed a ranking model composed of two separate neural networks trained jointly, one that matches query and document using local representation and another using learned distributed representation. In \cite{dehghani2017neural} input text was represented with bag-of-words and averaged bag-of-embeddings, and it was proposed three point-wise and pair-wise models, trained using weekly supervised data.

Transformers \cite{vaswani2017attention} have moved natural language processing (NLP) field from manually handcrafting features to learning semantic and deeper text representations. The Deep Learning Track at TREC 2019 \cite{craswell2020overview} hosted an extensive assessment of retrieval models after BERT \cite{devlin2018bert} was introduced by Google, and demonstrated the effectiveness of leveraging pre-trained transformers as ranking models \cite{Hambarde2023}.

In \cite{nogueira2019passage} BERT was adapted as a passage re-ranker. They proposed leveraging the \textit{next sentence prediction} training task to feed as input the concatenated query and passage separated by a special token and adding on top of the [CLS] output vector a single layer binary classification model. The authors later proposed in \cite{nogueira2019multi} a multi-stage retrieval pipeline composed by BM25 and two BERT ranking models: one trained with point-wise loss (monoBERT) and the other with a pair-wise loss (duoBERT). 

To deal with BERT limitation of 512 tokens, \cite{yang1903simple} proposed splitting documents into sentences for BERT usage. In \cite{qiao2019understanding} they experiment with bi-encoder and cross-encoder ranking models based on BERT, the latter being more effective due to the deeper interaction it provides. They also noticed BERT is more effective when working with semantically close query-document pairs compared to using click data for training.

Since then, a number of \textit{cross-encoder} models based on Transformers have been released from the community \footnote{\url{https://huggingface.co/cross-encoder}} \footnote{\url{https://www.sbert.net/docs/cross\_encoder/pretrained\_models.html}}. We discuss and compare some of the most accurate ranking models available publicly in the next section.

\section{Benchmarking ranking models for Q\&A text retrieval}
\label{sec:benchmarking_section}

Benchmarking models in terms of accuracy is important to support decision on which models should be used in production pipelines or fine-tuned for domain adaptation.

In this section, we evaluate text retrieval pipelines composed by different embedding and ranking models. To ensure the usefulness of this benchmark for companies, we evaluate the pipelines using three commercially usable embedding models and top-performing ranking models.

We emphasize our evaluation on Question-Answering (Q\&A) datasets, as that is a popular application of RAG systems. 

\subsection{Retrieval models}
\label{sec:retrieval}
There are many embedding models publicly available for the community, whose accuracy for multiple tasks can be found at the MTEB leaderboard\footnote{\url{https://huggingface.co/spaces/mteb/leaderboard}}. Most of those models have being trained on research-only datasets like MS-MARCO and cannot be used commercially. 

We evaluate embedding models that can be used for industry text retrieval applications. We select for experiments three embedding models for candidate retrieval, as the emphasis of our experiments is  evaluating ranking models:

\begin{itemize}
    \item \textit{Snowflake/snowflake-arctic-embed-l} \cite{merrick2024arctic} \footnote{\url{https://huggingface.co/Snowflake/snowflake-arctic-embed-l}} \footnote{\url{https://build.nvidia.com/snowflake/arctic-embed-l}} (335M params) - Embedding model based on BERT-large, trained with contrastive learning for two training rounds, with in-batch negative samples and hard negatives.
    \item \textit{nvidia/nv-embedqa-e5-v5} \footnote{\url{https://build.nvidia.com/nvidia/nv-embedqa-e5-v5}} (335M params) - Embedding model based on \textit{e5-large-unsupervised} trained for multiple rounds on supervised data with contrastive learning, and both in-batch negatives and hard-negatives.
    \item \textit{nvidia/nv-embedqa-mistral-7b-v2} \footnote{\url{https://build.nvidia.com/nvidia/nv-embedqa-mistral-7b-v2}} (7.24B params)- Large embedding model based on Mistral 7B v0.1 \cite{jiang2023mistral} \footnote{\url{https://huggingface.co/mistralai/Mistral-7B-v0.1}}, modified to use bi-directional attention and average pooling as done in NV-Embed \cite{lee2024nv} and NV-Retriever \cite{moreira2024nv}.
\end{itemize}

\subsection{Ranking models}
\label{sec:ranking}
We describe here the ranking models we evaluated in this investigation, including recent ranking models that perform high on retrieval benchmarks according to their reports. From those, the only reranking models that can be used for commercial purposes from their licences and train data are \textit{NV-RerankQA-Mistral-4B-v3}, which we introduce in this paper, and \textit{mixedbread-ai/mxbai-rerank-large-v1}.

\begin{itemize}
    \item \textit{ms-marco-MiniLM-L-12-v2} \footnote{\url{https://huggingface.co/cross-encoder/ms-marco-MiniLM-L-12-v2}} (33M params) - fine-tuned on top of the MiniLMv2 model \cite{wang2020minilmv2} with \textit{SentenceTransformers} package\footnote{\url{https://www.sbert.net/}} on the MS MARCO passage ranking dataset \cite{bajaj2016ms}.

    \item \textit{jina-reranker-v2-base-multilingual}
\footnote{\url{https://huggingface.co/jinaai/jina-reranker-v2-base-multilingual}} (278M params) - A multi-lingual ranking model finetuned from XLM-RoBERTa \cite{conneau2019unsupervised}

\item \textit{mixedbread-ai/mxbai-rerank-large-v1}
\footnote{\url{https://huggingface.co/mixedbread-ai/mxbai-rerank-large-v1}} (435M params) - Largest re-ranker model provided by Mixedbread

    \item \textit{bge-reranker-v2-m3} \footnote{\url{https://huggingface.co/BAAI/bge-reranker-v2-m3}} (568M params) - A multi-lingual ranking model fine-tuned from \textit{BGE M3-Embedding} \cite{chen2024bge} with FlagEmbedding package\footnote{\url{https://github.com/FlagOpen/FlagEmbedding}}

    \item \textit{NV-RerankQA-Mistral-4B-v3} \footnote{\url{https://build.nvidia.com/explore/retrieval\#nv-rerankqa-mistral-4b-v3}} (4B params) - Large and powerful re-ranker that takes as base model a pruned version of Mistral 7B and is fine-tuned with a blend of supervised data with contrastive learning. It is fully described in Section~\ref{sec:mistral_4b}.
\end{itemize}

\subsection{Evaluation setup}
The evaluation setup mimics the typical text retrieval indexing and querying pipelines, as previously illustrated in Figure~\ref{fig:pipelines}.

The indexing pipeline takes place first, where a text corpus is chunked into smaller passages. For our evaluation, we use datasets from BEIR \cite{thakur2021beir} datasets, which are already chunked and truncated to max 512 tokens. The chunked passages are embedded using an embedding model and stored in a vector index / database.
The querying pipeline then takes place for providing for each query a list with ranked passages for retrieval metrics computation (NDCG@10). In detail, the question is embedded  and it is performed a vector search (e.g. using exact or Approximate Nearest Neighbour (ANN) algorithm) on the vector index, returning the top-k most relevant passages for the question. Finally, the top-k (set to 100 in our evaluation experiments) passages are re-ranked with a ranking model to generate the final ordered list.

We perform the evaluation on the three Question-Answering datasets from BEIR \cite{thakur2021beir} retrieval benchmark: Natural Questions (NQ) \cite{kwiatkowski2019natural}, HotpotQA \cite{yang2018hotpotqa} and FiQA \cite{maia201818}.

\subsection{Benchmark results}
\label{ref:sec_results}
In this section, we provide the benchmark results of text retrieval pipelines with different embedding and ranking models.

In Table~\ref{tab:benchmark}, we compare pipelines with three commercially usable embedding models (Section~\ref{sec:retrieval}) and their combination with a number of ranking models (Section~\ref{sec:ranking}). Retrieval accuracy is measured with NDCG@10 for Q\&A BEIR datasets.

\begin{table}[ht]
\caption{Evaluation (NDCG@10) of multi-stage text retrieval pipelines with different embedding and ranking models on text Q\&A datasets from BEIR}
\footnotesize
\begin{tabular}{p{4.4cm}l|p{0.5cm}p{0.9cm}l}
\textbf{Reranker model}                    & \textbf{Avg.}   & \textbf{NQ}     & \textbf{HotpotQA} & \textbf{FiQA}   \\ \hline
\textbf{\textit{Embedding:} snowflake-arctic-embed-l}                        & 0.6100 & 0.6311 & 0.7518   & 0.4471 \\ \hline
+ ms-marco-MiniLM-L-12-v2            & 0.5771 & 0.5876 & 0.7586   & 0.3850 \\
+ mxbai-rerank-large-v1              & 0.6077 & 0.6433 & 0.7401   & 0.4396 \\
+ jina-reranker-v2-base-multilingual & 0.6481 & 0.6768 & 0.8165   & 0.4511 \\
+ bge-reranker-v2-m3                 & 0.6585 & 0.6965 & 0.8458   & 0.4332 \\

+ NV-RerankQA-Mistral-4B-v3          & \textbf{0.7529} & \textbf{0.7788} & \textbf{0.8726}   & \textbf{0.6073} \\ \hline \hline

\textbf{\textit{Embedding:} NV-EmbedQA-e5-v5}                        & 0.6083 & 0.6380 & 0.7160   & 0.4710 \\ \hline
+ ms-marco-MiniLM-L-12-v2            & 0.5785 & 0.5909 & 0.7458   & 0.3988 \\
+ mxbai-rerank-large-v1              & 0.6077 & 0.6450 & 0.7279   & 0.4502 \\
+ jina-reranker-v2-base-multilingual & 0.6454 & 0.6780 & 0.7996   & 0.4585 \\
+ bge-reranker-v2-m3                 & 0.6584 & 0.6974 & 0.8272   & 0.4506 \\
+ NV-RerankQA-Mistral-4B-v3          & \textbf{0.7486} & \textbf{0.7785} & \textbf{0.8470}   & \textbf{0.6203} \\ \hline \hline

\textbf{\textit{Embedding:}   NV-EmbedQA-Mistral7B-v2}                        & 0.7173 & 0.7216 & 0.8109   & 0.6194 \\ \hline
+ ms-marco-MiniLM-L-12-v2            & 0.5875 & 0.5945 & 0.7641   & 0.4039 \\
+ mxbai-rerank-large-v1              & 0.6133 & 0.6439 &   0.7436   & 0.4523 \\
+ jina-reranker-v2-base-multilingual & 0.6590 & 0.6819 & 0.8262 &   0.4689                 \\
+ bge-reranker-v2-m3                 & 0.6734 & 0.7028 & 0.8635   & 0.4539 \\
+ NV-RerankQA-Mistral-4B-v3          & \textbf{0.7694} & \textbf{0.7830} & \textbf{0.8904}   & \textbf{0.6350} \\ \hline

\end{tabular}
\label{tab:benchmark}
\end{table}

We can clearly observe that for smaller embedding models like \textit{snowflake-arctic-embed-l} and \textit{NV-EmbedQA-e5-v5} (335M params), all cross-encoders (except for the small \textit{ms-marco-MiniLM-L-12-v2}) improve considerably the ranking accuracy compared to the retriever. On the other hand, for the larger \textit{NV-EmbedQA-Mistral7B-v2} embedding model, only the large \textit{NV-RerankQA-Mistral-4B-v3} reranker is able to improve its accuracy.

The \textit{NV-RerankQA-Mistral-4B-v3} reranker provides the highest ranking accuracy for all datasets by a large margin (+14\% compared to the second best reranker: \textit{bge-reranker-v2-m3}). That demonstrates the effectiveness of our adaptation of Mistral 7B as a cross-encoder.

\subsection{A small note about model licensing}

For training \textit{NV-RerankQA-Mistral-4B-v3} we have selected only public datasets whose license allows their usage for industry applications. Some other models are released with permissive licenses like Apache 2.0 or MIT, but we do not know which datasets they were trained on or whether they got a special license to use research-only datasets like MS-Marco, for example. Every company should check with its legal team on model licensing for commercial usage.

\section{Fine-tuning a state-of-the-art ranking model: \textit{NV-RerankQA-Mistral-4B-v3}}
\label{sec:mistral_4b}
We introduce in this paper the state-of-the-art \textit{NV-RerankQA-Mistral-4B-v3}, that performs best in our benchmark on text retrieval for Q\&A (Section~\ref{ref:sec_results}).

Mistral 7B  \cite{jiang2023mistral} decoder model has been successfully adopted as embedding models for retrieval when repurposed and fine-tuned with contrastive learning\cite{wang2023improving, lee2024nv, moreira2024nv}. 

In this work, we adapted Mistral 7B v0.1 \cite{jiang2023mistral} as a ranking model. In order to reduce the number of parameters from the base model, thus its inference compute and memory requirements, we prune it by keeping only the bottom 16 layers out of its 32 layers\footnote{We also tried pruning different number of top layers from Mistral 7B model, the more layers we remove the lower the accuracy. We found out that keeping bottom 16 layers provides a good trade-off between accuracy penalty (-1\%) and model size reduction (-50\% \# parameters) compared to the original 32-layer model.}. We also modify its self-attention mechanism from uni-directional (causal) to bi-directional, so that for each token it is possible to attend to other tokens in both right and left sides, as that has shown to improve accuracy for Mistral-based embedding models \cite{lee2024nv, moreira2024nv}.

We feed as input to the model the tokenized question and candidate passage pair, concatenated and separated by a special token. We perform average pooling on the outputs of last Transformer layer and add a  feed-forward layer on top that outputs a single-unit with the likelihood of a given passage being relevant to a question.

\begin{figure}[ht]
    \centering
    \includegraphics[width=0.9\linewidth]{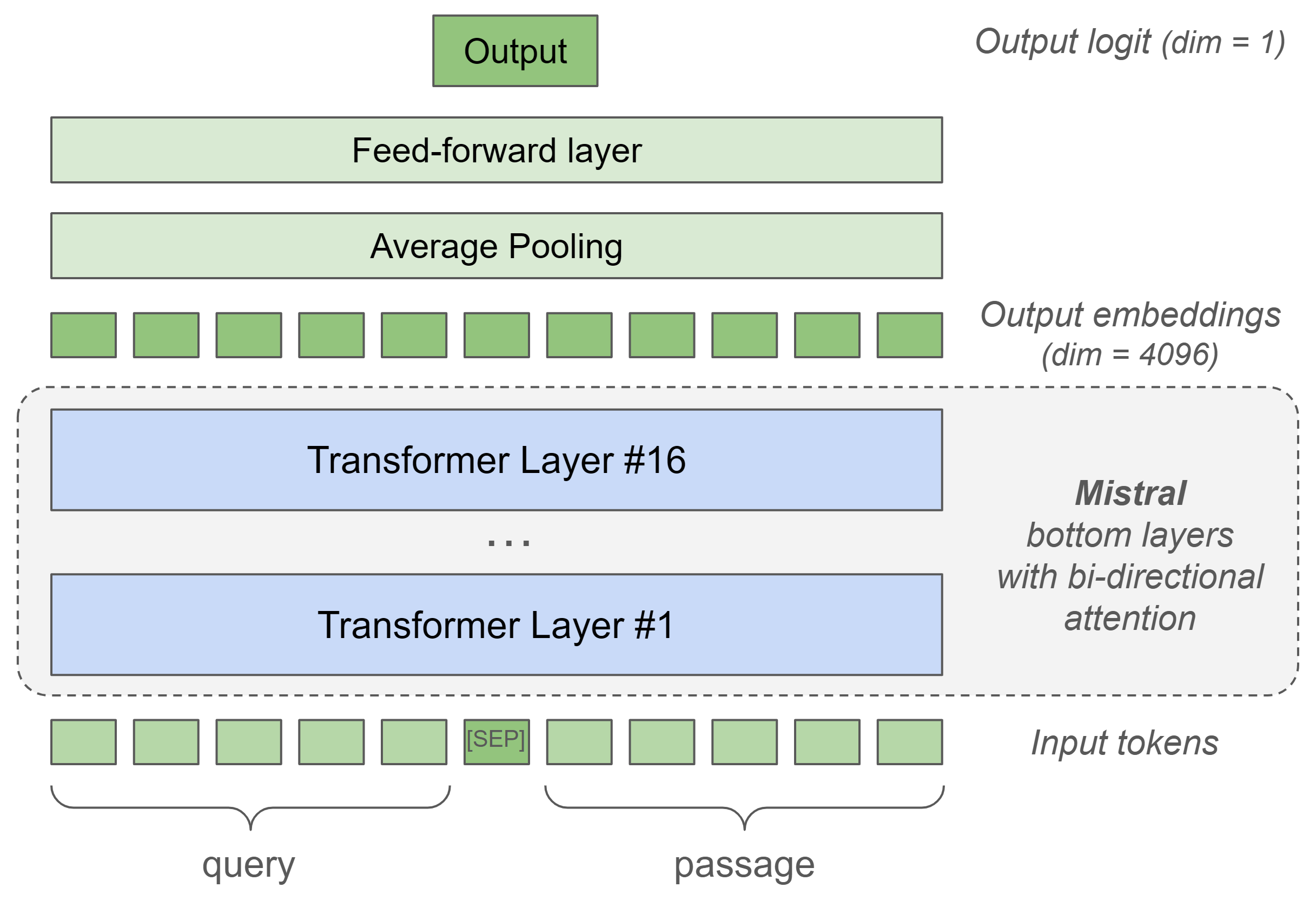}
    \caption{Architecture of \textit{NV-RerankQA-Mistral-4B-v3} cross-encoder, pruned and adapted from Mistral 7B}
    \label{fig:mistral_4b}
\end{figure}

Cross-encoder ranking models are binary classifiers that discriminate between positive and negative passages. They typically are trained with the binary cross-entropy loss as in Equation~\ref{eq:bce}, where $ p = \phi(q,d) $ is the model predicted likelihood of the passage $ d $ being relevant to query $q$. 

\begin{equation}
    L = -{(y\log(p) + (1 - y)\log(1 - p))}
\label{eq:bce}
\end{equation}

Instead, for \textit{NV-RerankQA-Mistral-4B-v3} we follow \cite{wang2022simlm} and train the reranker with contrastive learning over the positive and its negative candidates scores using the list-wise InfoNCE loss \cite{oord2018representation}, shown in Equation~\ref{eq:infonce}, where $ d^+ $ is a positive relevant passage, $ d^-$ is one of the $ N $ negative passages and $ \tau $ is the temperature parameter.

\begin{equation}
    L = -\text{log} \frac{ \text{exp}(\phi(q,d^+)/\tau)) }{ \text{exp}(\phi(q,d^+)/\tau)) + \sum_{i=1}^{N}  \text{exp}(\phi(q,d^-)/\tau))}
\label{eq:infonce}
\end{equation}

The negative candidates used for contrastive learning are mined from the corpus in the data pre-processing stage by using a teacher embedding model. We use the \textit{TopK-PercPos} hard-negative mining method introduced in \cite{moreira2024nv}, configured with maximum negative score threshold as 95\% of the positive scores to remove potential false negatives.

We present in Section~\ref{sec:ablation} an ablation study on fine-tuning ranking models, with some experiments focused on our choices for the loss and self-attention mechanism for \textit{NV-RerankQA-Mistral-4B-v3}.

\section{Ablation study on fine-tuning ranking models}
\label{sec:ablation}

In this section, we present an ablation study on fine-tuning and comparing different base models as rerankers. For Mistral base model, we also compare choices of self-attention mechanism (unidirectional vs bi-directional) and training losses (binary vs categorical cross-entropy).

For broader comparison, we evaluate the ranking models with the three different embedding models described in Section~\ref{sec:retrieval}.

\subsection{Model size matters}
Model size is an important aspect for trading-off model accuracy and inference throughput. For this section we fine-tune and compare three base models with different sizes as ranking models: \textit{MiniLM-L12-H384-uncased} \cite{wang2020minilm} (33M params) \footnote{\url{https://huggingface.co/microsoft/MiniLM-L12-H384-uncased}}, \textit{deberta-v3-large} \cite{he2021debertav3} \footnote{\url{https://huggingface.co/microsoft/deberta-v3-large}} (435M) and \textit{Mistral 4B} (4B params), the latter pruned and modified from Mistral 7B v0.1 \cite{jiang2023mistral} as described in Section~\ref{sec:mistral_4b}. 

Those ranking models are all fine-tuned with the same compute budget (max 4 hours of training in a single server with 8x A100 GPUs) and same train set.

In Table~\ref{tab:model_size}, we present the comparison of fine-tuning those different ranking models, with the same train set and compute budget. Although the pre-training of those base models is different, it is possible to observe a pattern where larger ranking models provide higher retrieval accuracy. 

The most accurate model is based on \textit{Mistral 4B}, but \textit{deberta-v3-large} is surprisingly accurate for its smaller number of parameters,  and is a good candidate architecture for deploying as a cross-encoder, as we discuss in Section~\ref{sec:deployment}.

\begin{table}[ht]
\caption{Comparing multi-stage text retrieval pipelines with different-sized ranking models, fine-tuned with the same train set and compute budget. Metric is NDCG@10.}
\footnotesize
\begin{tabular}{p{4.4cm}l|p{0.5cm}p{0.9cm}l}
\textbf{Reranker model}                    & \textbf{Avg.}   & \textbf{NQ}     & \textbf{HotpotQA} & \textbf{FiQA}   \\ \hline

\textbf{\textit{Embedding:} snowflake-arctic-embed-l}                              & 0.6100 & 0.6311 & 0.7518   & 0.4471 \\ \hline
+ MiniLM-L12-H384-uncased                                & 0.6227 & 0.6436 & 0.8128   & 0.4118 \\
+ deberta-v3-large                                & 0.7277 & 0.7452 & 0.8548   & 0.5832 \\
+ Mistral 4B                                & \textbf{0.7414} & \textbf{0.7690} & \textbf{0.8681}   & \textbf{0.5872} \\
\hline  \hline
\textbf{\textit{Embedding:} NV-EmbedQA-e5-v5}                               & 0.6083 & 0.6380 & 0.7160   & 0.4710 \\ \hline
+ MiniLM-L12-H384-uncased                                & 0.6213 & 0.6438 & 0.7954   & 0.4248 \\
+ deberta-v3-large                                & 0.7150 & 0.7441 & 0.8319   & 0.5689 \\
+ Mistral 4B                                & \textbf{0.7366} & \textbf{0.7689} & \textbf{0.8423}   & \textbf{0.5987} \\
\hline \hline
\textbf{\textit{Embedding:}   NV-EmbedQA-Mistral7B-v2}                              & 0.7173 & 0.7216 & 0.8109   & 0.6194 \\ \hline
+ MiniLM-L12-H384-uncased                                & 0.6355 & 0.6484 & 0.8269   & 0.4312 \\
+ deberta-v3-large                                & 0.7413 & 0.7486 & 0.8700   & 0.6055 \\
+ Mistral 4B                                & \textbf{0.7575} & \textbf{0.7717} & \textbf{0.8857}   & \textbf{0.6152} \\ \hline
\end{tabular}
\label{tab:model_size}
\end{table}

\subsection{Causal vs Bi-directional Attention mechanism}
In Section~\ref{sec:mistral_4b} we describe that for our adapted \textit{Mistral 4B} we modified the standard self-attention mechanism of Mistral from uni-directional (causal) to bi-directional attention.

We compare the accuracy with those two self-attention mechanisms in Table~\ref{tab:attention}, both using average pooling, and demonstrate the effectiveness of bi-directional attention for allowing deeper interaction among input query and passage tokens.

\begin{table}[ht]
\caption{Comparing NDCG@10 of Mistral 4B reranker fine-tuned with different attention mechanisms.}
\footnotesize
\begin{tabular}{p{4.4cm}l|p{0.5cm}p{0.9cm}l}
\textbf{Reranker model}                    & \textbf{Avg.}   & \textbf{NQ}     & \textbf{HotpotQA} & \textbf{FiQA}   \\ \hline

\textbf{\textit{Embedding:} snowflake-arctic-embed-l}                              & 0.6100 & 0.6311 & 0.7518   & 0.4471 \\ \hline
+ Mistral 4B (unidirectional attention)     & 0.7312 & 0.7663 & 0.8612   & 0.5660 \\
+ Mistral 4B (bidirectional attention)      & \textbf{0.7414} & \textbf{0.7690} & \textbf{0.8681}   & \textbf{0.5872} \\
\hline  \hline
\textbf{\textit{Embedding:} NV-EmbedQA-e5-v5}                               & 0.6083 & 0.6380 & 0.7160   & 0.4710 \\ \hline
+ Mistral 4B (unidirectional attention)     & 0.7264 & 0.7655 & 0.8372   & 0.5766 \\
+ Mistral 4B (bidirectional attention)      & \textbf{0.7366} & \textbf{0.7689} & \textbf{0.8423}   & \textbf{0.5987} \\
\hline \hline
\textbf{\textit{Embedding:}   NV-EmbedQA-Mistral7B-v2}                              & 0.7173 & 0.7216 & 0.8109   & 0.6194 \\ \hline
+ Mistral 4B (unidirectional attention)     & 0.7464 & 0.7690 & 0.8781   & 0.5920 \\
+ Mistral 4B (bidirectional attention)      & \textbf{0.7575} & \textbf{0.7717} & \textbf{0.8857}   & \textbf{0.6152} \\
\hline
\end{tabular}
\label{tab:attention}
\end{table}

\subsection{BCE vs InfoNCE Loss}
Cross-encoders are typically trained with the point-wise Binary Cross-Entropy (BCE) loss (Equation~\ref{eq:bce}), as we discussed in Section~\ref{sec:mistral_4b}.

On the other hand, we fine-tune \textit{Mistral 4B} with the list-wise InfoNCE loss\cite{oord2018representation} (Equation~\ref{eq:infonce}) and contrastive learning.

We experiment with those two losses, both using the same sets of hard-negative passages mined from the corpus, as described in Section~\ref{sec:mistral_4b}.

In Table~\ref{tab:loss}, we can clearly observe the higher retrieval accuracy obtained when using InfoNCE,  a list-wise contrastive learning loss trained to maximize the relevance score of the question and positive passage pair, while minimizing the score for question and negative passage pairs.

\begin{table}[ht]
\caption{Comparing NDCG@10 of Mistral 4B reranker with bi-directional attention fine-tuned with different losses.}
\footnotesize
\begin{tabular}{p{4.4cm}l|p{0.5cm}p{0.9cm}l}
\textbf{Reranker model}                    & \textbf{Avg.}   & \textbf{NQ}     & \textbf{HotpotQA} & \textbf{FiQA}   \\ \hline

\textbf{\textit{Embedding:} snowflake-arctic-embed-l}                              & 0.6100 & 0.6311 & 0.7518   & 0.4471 \\ \hline
+ Mistral 4B (BCE loss) &  0.7230   & 0.7375 & 0.8609  &  0.5706   \\		
+ Mistral 4B (InfoNCE loss)      & \textbf{0.7414} & \textbf{0.7690} & \textbf{0.8681}   & \textbf{0.5872} \\
\hline  \hline
\textbf{\textit{Embedding:} NV-EmbedQA-e5-v5}                               & 0.6083 & 0.6380 & 0.7160   & 0.4710 \\ \hline
+ Mistral 4B (BCE loss)     & 0.7171    & 0.7368  & 0.8357 & 0.5786 \\
+ Mistral 4B (InfoNCE loss)      & \textbf{0.7366} & \textbf{0.7689} & \textbf{0.8423}   & \textbf{0.5987} \\
\hline \hline
\textbf{\textit{Embedding:}   NV-EmbedQA-Mistral7B-v2}                              & 0.7173 & 0.7216 & 0.8109   & 0.6194 \\ \hline
+ Mistral 4B (BCE loss)    & 0.7373 & 0.7394 &  0.8774  & 0.5949 \\
+ Mistral 4B (InfoNCE loss)      & \textbf{0.7575} & \textbf{0.7717} & \textbf{0.8857}   & \textbf{0.6152} \\
\hline
\end{tabular}
\label{tab:loss}
\end{table}

This ablation study explains our choices of using bi-directional attention and InfoNCE loss for fine-tuning \textit{NV-RerankQA-Mistral-4B-v3}. 

\section{Deployment trade-off considerations for text retrieval pipelines with ranking models}
\label{sec:deployment}

As we discussed before, the model sizes and usage of ranking models have implications on the performance of the deployed text retrieval pipelines and downstream systems that use it, such as RAG applications. Deploying the indexing pipeline to production using a large embedding model would be computationally expensive, especially if the document corpus is large. Furthermore, when we deploy query pipeline to production, it is critical that it can handle a large number of queries in a timely manner and scale on demand. 
In some cases, it might be possible to improve both the retrieval accuracy and indexing throughput by replacing a single-stage query pipeline of a large embedding model by a two-stage pipeline composed of a smaller embedding model and a ranking model.

We conduct performance experiments with our models optimized\footnote{You can find Nemo Retriever embedding and ranking models optimized for fast serving with TensorRT and Triton Inference Server in \url{https://docs.nvidia.com/nim/index.html\#nemo-retriever}} with TensorRT\footnote{\url{https://developer.nvidia.com/tensorrt}} and deployed on Triton Inference Server\footnote{\url{https://developer.nvidia.com/triton-inference-server}}. In Table~\ref{tab:embedding_perf}, we present the average query embedding latency and passages indexing throughput for two embedding models with different sizes. Although the time difference to embed a single query with those two models will not compromise the overall online retrieval latency, the indexing time of the chunked passages of the textual corpus will take 8.2x longer with the larger embedding model, which results in higher compute/cost requirements when re-indexing is needed (e.g. when corpus or embedding model changes).

\begin{table}[ht]
\caption{Average query embedding latency (batch size=1 question with 20 tokens) and passages indexing throughput (batch size=64 passages with 512 tokens), with models converted to FP-16 and TensorRT  deployed on Triton Inference Server on a single H100-HBM3-80GB GPU\tablefootnote{You can find more information on Nemo Retriever performance benchmarks with other GPUs / batch sizes for embedding models in \url{https://docs.nvidia.com/nim/nemo-retriever/text-embedding/latest/performance.html} and for ranking models in \url{https://docs.nvidia.com/nim/nemo-retriever/text-reranking/latest/performance.html}}}.
\footnotesize
\begin{tabular}{lp{2.0cm}p{2.3cm}}
\textbf{Embedding model} & \textbf{Query embedding latency} & \textbf{Passages indexing throughput} \\ \hline
NV-EmbedQA-E5-v5         & \multicolumn{1}{r}{5.1 ms}                        & \multicolumn{1}{r}{558.4 passages/sec}                \\
NV-EmbedQA-Mistral7B-v2  & \multicolumn{1}{r}{19.8 ms}                       & \multicolumn{1}{r}{68.7 passages/sec}                 \\ \hline
\end{tabular}
\label{tab:embedding_perf}
\end{table}

Thus, by using a two-stage pipeline with \textit{NV-EmbedQA-E5-v5} embedding and \textit{NV-RerankQA-Mistral-4B-v3} ranking models instead of a single-stage pipeline with \textit{NV-EmbedQA-Mistral7B-v2}, in addition to achieving higher retrieval accuracy (see Table~\ref{tab:benchmark}), we will also reduce the indexing time by 8.2x.

We also have to consider the latency the ranking model adds to the query pipeline. In our example, after the query is embedded with \textit{NV-EmbedQA-E5-v5} (in 5.1 ms), candidate relevant passages are retrieved from a vector database using ANN, and the top-40 passages will be provided to a  ranking model for final ranking. The 40 candidate passages would be scored on their relevancy with respect to the query by the \textit{NV-RerankQA-Mistral-4B-v3} model in total 266 ms on average, which might add a reasonable time to the overall query pipeline latency depending on the non-functional requirements for the system. That could be improved by distributing the ranking scoring requests of those 40 candidates among multiple GPUs / Triton Inference Server instances. Another option would be deploying a model with a good trade-off between size/latency and accuracy, like \textit{deberta-v3-large}\footnote{We are actually building a ranking model based on \textit{deberta-v3-large} and plan to release it soon.}, which is much smaller (435M) than \textit{NV-RerankQA-Mistral-4B-v3} (4B) and might provide a good ranking accuracy as discussed in the ablation study on Table~\ref{tab:model_size}.

In summary, the decision of which models to include in a text retrieval pipeline should consider the business requirements for retrieval accuracy and system requirements on indexing throughput and serving latency. For example, if the text corpus to index is huge, probably indexing throughput will be the bottleneck and smaller embedding models should be used. On the other hand, if serving latency requirements are very strict, a fast query pipeline is more critical, and large ranking models should be avoided.

\section{Conclusion}

In this paper, we provide a comprehensive evaluation of multi-stage text retrieval pipelines for Question-Answering, a common use case for RAG applications. 
The evaluated pipelines composed of commercially viable embedding models and state-of-the-art ranking models. We introduce the \textit{NV-RerankQA-Mistral-4B-v3}, that provides the best ranking accuracy by a large margin in our benchmark and is commercially usable for industrial applications.

We describe how we adapted the decoder-only \textit{Mistral 7B} and fine-tuned it as a cross-encoder, pruning the base model and modifying its attention and pooling mechanism to build the \textit{NV-RerankQA-Mistral-4B-v3}.

We also provide an ablation study comparing the fine-tuning of different-sized base models as cross-encoders, and highlight the relationship between their number of parameters and ranking accuracy. We also compare the benefits of leveraging bi-directional attention and InfoNCE loss for training a Mistral cross-encoder.

Finally, we discussed important deployment considerations for real-world text retrieval systems, with respect to trading-off model size, retrieval accuracy and systems requirements like serving latency and indexing throughput.

\bibliographystyle{ACM-Reference-Format}
\bibliography{manuscript}


\begin{thebibliography}{38}

%%% the Web address.

\ifx \showCODEN    \undefined \def \showCODEN     #1{\unskip}     \fi
\ifx \showDOI      \undefined \def \showDOI       #1{#1}\fi
\ifx \showISBNx    \undefined \def \showISBNx     #1{\unskip}     \fi
\ifx \showISBNxiii \undefined \def \showISBNxiii  #1{\unskip}     \fi
\ifx \showISSN     \undefined \def \showISSN      #1{\unskip}     \fi
\ifx \showLCCN     \undefined \def \showLCCN      #1{\unskip}     \fi
\ifx \shownote     \undefined \def \shownote      #1{#1}          \fi
\ifx \showarticletitle \undefined \def \showarticletitle #1{#1}   \fi
\ifx \showURL      \undefined \def \showURL       {\relax}        \fi
% invisible to TeX
\providecommand\bibfield[2]{#2}
\providecommand\bibinfo[2]{#2}
\providecommand\natexlab[1]{#1}
\providecommand\showeprint[2][]{arXiv:#2}

\bibitem[Bajaj et~al\mbox{.}(2016)]%
        {bajaj2016ms}
\bibfield{author}{\bibinfo{person}{Payal Bajaj}, \bibinfo{person}{Daniel Campos}, \bibinfo{person}{Nick Craswell}, \bibinfo{person}{Li Deng}, \bibinfo{person}{Jianfeng Gao}, \bibinfo{person}{Xiaodong Liu}, \bibinfo{person}{Rangan Majumder}, \bibinfo{person}{Andrew McNamara}, \bibinfo{person}{Bhaskar Mitra}, \bibinfo{person}{Tri Nguyen}, {et~al\mbox{.}}} \bibinfo{year}{2016}\natexlab{}.
\newblock \showarticletitle{Ms marco: A human generated machine reading comprehension dataset}.
\newblock \bibinfo{journal}{\emph{arXiv preprint arXiv:1611.09268}} (\bibinfo{year}{2016}).
\newblock


\bibitem[Chen et~al\mbox{.}(2024)]%
        {chen2024bge}
\bibfield{author}{\bibinfo{person}{Jianlv Chen}, \bibinfo{person}{Shitao Xiao}, \bibinfo{person}{Peitian Zhang}, \bibinfo{person}{Kun Luo}, \bibinfo{person}{Defu Lian}, {and} \bibinfo{person}{Zheng Liu}.} \bibinfo{year}{2024}\natexlab{}.
\newblock \showarticletitle{Bge m3-embedding: Multi-lingual, multi-functionality, multi-granularity text embeddings through self-knowledge distillation}.
\newblock \bibinfo{journal}{\emph{arXiv preprint arXiv:2402.03216}} (\bibinfo{year}{2024}).
\newblock


\bibitem[Chen et~al\mbox{.}(2020)]%
        {chen2020simple}
\bibfield{author}{\bibinfo{person}{Ting Chen}, \bibinfo{person}{Simon Kornblith}, \bibinfo{person}{Mohammad Norouzi}, {and} \bibinfo{person}{Geoffrey Hinton}.} \bibinfo{year}{2020}\natexlab{}.
\newblock \showarticletitle{A simple framework for contrastive learning of visual representations}. In \bibinfo{booktitle}{\emph{International conference on machine learning}}. PMLR, \bibinfo{pages}{1597--1607}.
\newblock


\bibitem[Conneau et~al\mbox{.}(2019)]%
        {conneau2019unsupervised}
\bibfield{author}{\bibinfo{person}{Alexis Conneau}, \bibinfo{person}{Kartikay Khandelwal}, \bibinfo{person}{Naman Goyal}, \bibinfo{person}{Vishrav Chaudhary}, \bibinfo{person}{Guillaume Wenzek}, \bibinfo{person}{Francisco Guzm{\'a}n}, \bibinfo{person}{Edouard Grave}, \bibinfo{person}{Myle Ott}, \bibinfo{person}{Luke Zettlemoyer}, {and} \bibinfo{person}{Veselin Stoyanov}.} \bibinfo{year}{2019}\natexlab{}.
\newblock \showarticletitle{Unsupervised cross-lingual representation learning at scale}.
\newblock \bibinfo{journal}{\emph{arXiv preprint arXiv:1911.02116}} (\bibinfo{year}{2019}).
\newblock


\bibitem[Craswell et~al\mbox{.}(2020)]%
        {craswell2020overview}
\bibfield{author}{\bibinfo{person}{Nick Craswell}, \bibinfo{person}{Bhaskar Mitra}, \bibinfo{person}{Emine Yilmaz}, \bibinfo{person}{Daniel Campos}, {and} \bibinfo{person}{Ellen~M Voorhees}.} \bibinfo{year}{2020}\natexlab{}.
\newblock \showarticletitle{Overview of the TREC 2019 deep learning track}.
\newblock \bibinfo{journal}{\emph{arXiv preprint arXiv:2003.07820}} (\bibinfo{year}{2020}).
\newblock


\bibitem[Dehghani et~al\mbox{.}(2017)]%
        {dehghani2017neural}
\bibfield{author}{\bibinfo{person}{Mostafa Dehghani}, \bibinfo{person}{Hamed Zamani}, \bibinfo{person}{Aliaksei Severyn}, \bibinfo{person}{Jaap Kamps}, {and} \bibinfo{person}{W~Bruce Croft}.} \bibinfo{year}{2017}\natexlab{}.
\newblock \showarticletitle{Neural ranking models with weak supervision}. In \bibinfo{booktitle}{\emph{Proceedings of the 40th international ACM SIGIR conference on research and development in information retrieval}}. \bibinfo{pages}{65--74}.
\newblock


\bibitem[Devlin et~al\mbox{.}(2018)]%
        {devlin2018bert}
\bibfield{author}{\bibinfo{person}{Jacob Devlin}, \bibinfo{person}{Ming-Wei Chang}, \bibinfo{person}{Kenton Lee}, {and} \bibinfo{person}{Kristina Toutanova}.} \bibinfo{year}{2018}\natexlab{}.
\newblock \showarticletitle{Bert: Pre-training of deep bidirectional transformers for language understanding}.
\newblock \bibinfo{journal}{\emph{arXiv preprint arXiv:1810.04805}} (\bibinfo{year}{2018}).
\newblock


\bibitem[Guo et~al\mbox{.}(2016)]%
        {guo2016deep}
\bibfield{author}{\bibinfo{person}{Jiafeng Guo}, \bibinfo{person}{Yixing Fan}, \bibinfo{person}{Qingyao Ai}, {and} \bibinfo{person}{W~Bruce Croft}.} \bibinfo{year}{2016}\natexlab{}.
\newblock \showarticletitle{A deep relevance matching model for ad-hoc retrieval}. In \bibinfo{booktitle}{\emph{Proceedings of the 25th ACM international on conference on information and knowledge management}}. \bibinfo{pages}{55--64}.
\newblock


\bibitem[Hambarde and Proença(2023)]%
        {Hambarde2023}
\bibfield{author}{\bibinfo{person}{Kailash~A. Hambarde} {and} \bibinfo{person}{Hugo Proença}.} \bibinfo{year}{2023}\natexlab{}.
\newblock \showarticletitle{Information Retrieval: Recent Advances and Beyond}.
\newblock \bibinfo{journal}{\emph{IEEE Access}}  \bibinfo{volume}{11} (\bibinfo{year}{2023}), \bibinfo{pages}{76581--76604}.
\newblock
\urldef\tempurl%
\url{https://doi.org/10.1109/ACCESS.2023.3295776}
\showDOI{\tempurl}


\bibitem[He et~al\mbox{.}(2021)]%
        {he2021debertav3}
\bibfield{author}{\bibinfo{person}{Pengcheng He}, \bibinfo{person}{Jianfeng Gao}, {and} \bibinfo{person}{Weizhu Chen}.} \bibinfo{year}{2021}\natexlab{}.
\newblock \showarticletitle{Debertav3: Improving deberta using electra-style pre-training with gradient-disentangled embedding sharing}.
\newblock \bibinfo{journal}{\emph{arXiv preprint arXiv:2111.09543}} (\bibinfo{year}{2021}).
\newblock


\bibitem[Jiang et~al\mbox{.}(2023)]%
        {jiang2023mistral}
\bibfield{author}{\bibinfo{person}{Albert~Q Jiang}, \bibinfo{person}{Alexandre Sablayrolles}, \bibinfo{person}{Arthur Mensch}, \bibinfo{person}{Chris Bamford}, \bibinfo{person}{Devendra~Singh Chaplot}, \bibinfo{person}{Diego de~las Casas}, \bibinfo{person}{Florian Bressand}, \bibinfo{person}{Gianna Lengyel}, \bibinfo{person}{Guillaume Lample}, \bibinfo{person}{Lucile Saulnier}, {et~al\mbox{.}}} \bibinfo{year}{2023}\natexlab{}.
\newblock \showarticletitle{Mistral 7B}.
\newblock \bibinfo{journal}{\emph{arXiv preprint arXiv:2310.06825}} (\bibinfo{year}{2023}).
\newblock


\bibitem[Karpukhin et~al\mbox{.}(2020)]%
        {karpukhin2020dense}
\bibfield{author}{\bibinfo{person}{Vladimir Karpukhin}, \bibinfo{person}{Barlas O{\u{g}}uz}, \bibinfo{person}{Sewon Min}, \bibinfo{person}{Patrick Lewis}, \bibinfo{person}{Ledell Wu}, \bibinfo{person}{Sergey Edunov}, \bibinfo{person}{Danqi Chen}, {and} \bibinfo{person}{Wen-tau Yih}.} \bibinfo{year}{2020}\natexlab{}.
\newblock \showarticletitle{Dense passage retrieval for open-domain question answering}.
\newblock \bibinfo{journal}{\emph{arXiv preprint arXiv:2004.04906}} (\bibinfo{year}{2020}).
\newblock


\bibitem[Kwiatkowski et~al\mbox{.}(2019)]%
        {kwiatkowski2019natural}
\bibfield{author}{\bibinfo{person}{Tom Kwiatkowski}, \bibinfo{person}{Jennimaria Palomaki}, \bibinfo{person}{Olivia Redfield}, \bibinfo{person}{Michael Collins}, \bibinfo{person}{Ankur Parikh}, \bibinfo{person}{Chris Alberti}, \bibinfo{person}{Danielle Epstein}, \bibinfo{person}{Illia Polosukhin}, \bibinfo{person}{Jacob Devlin}, \bibinfo{person}{Kenton Lee}, {et~al\mbox{.}}} \bibinfo{year}{2019}\natexlab{}.
\newblock \showarticletitle{Natural questions: a benchmark for question answering research}.
\newblock \bibinfo{journal}{\emph{Transactions of the Association for Computational Linguistics}}  \bibinfo{volume}{7} (\bibinfo{year}{2019}), \bibinfo{pages}{453--466}.
\newblock


\bibitem[Lee et~al\mbox{.}(2024)]%
        {lee2024nv}
\bibfield{author}{\bibinfo{person}{Chankyu Lee}, \bibinfo{person}{Rajarshi Roy}, \bibinfo{person}{Mengyao Xu}, \bibinfo{person}{Jonathan Raiman}, \bibinfo{person}{Mohammad Shoeybi}, \bibinfo{person}{Bryan Catanzaro}, {and} \bibinfo{person}{Wei Ping}.} \bibinfo{year}{2024}\natexlab{}.
\newblock \showarticletitle{NV-Embed: Improved Techniques for Training LLMs as Generalist Embedding Models}.
\newblock \bibinfo{journal}{\emph{arXiv preprint arXiv:2405.17428}} (\bibinfo{year}{2024}).
\newblock


\bibitem[Lewis et~al\mbox{.}(2020)]%
        {lewis2020retrieval}
\bibfield{author}{\bibinfo{person}{Patrick Lewis}, \bibinfo{person}{Ethan Perez}, \bibinfo{person}{Aleksandra Piktus}, \bibinfo{person}{Fabio Petroni}, \bibinfo{person}{Vladimir Karpukhin}, \bibinfo{person}{Naman Goyal}, \bibinfo{person}{Heinrich K{\"u}ttler}, \bibinfo{person}{Mike Lewis}, \bibinfo{person}{Wen-tau Yih}, \bibinfo{person}{Tim Rockt{\"a}schel}, {et~al\mbox{.}}} \bibinfo{year}{2020}\natexlab{}.
\newblock \showarticletitle{Retrieval-augmented generation for knowledge-intensive nlp tasks}.
\newblock \bibinfo{journal}{\emph{Advances in Neural Information Processing Systems}}  \bibinfo{volume}{33} (\bibinfo{year}{2020}), \bibinfo{pages}{9459--9474}.
\newblock


\bibitem[Liu et~al\mbox{.}(2009)]%
        {liu2009learning}
\bibfield{author}{\bibinfo{person}{Tie-Yan Liu} {et~al\mbox{.}}} \bibinfo{year}{2009}\natexlab{}.
\newblock \showarticletitle{Learning to rank for information retrieval}.
\newblock \bibinfo{journal}{\emph{Foundations and Trends{\textregistered} in Information Retrieval}} \bibinfo{volume}{3}, \bibinfo{number}{3} (\bibinfo{year}{2009}), \bibinfo{pages}{225--331}.
\newblock


\bibitem[Lu and Li(2013)]%
        {lu2013deep}
\bibfield{author}{\bibinfo{person}{Zhengdong Lu} {and} \bibinfo{person}{Hang Li}.} \bibinfo{year}{2013}\natexlab{}.
\newblock \showarticletitle{A deep architecture for matching short texts}.
\newblock \bibinfo{journal}{\emph{Advances in neural information processing systems}}  \bibinfo{volume}{26} (\bibinfo{year}{2013}).
\newblock


\bibitem[Maia et~al\mbox{.}(2018)]%
        {maia201818}
\bibfield{author}{\bibinfo{person}{Macedo Maia}, \bibinfo{person}{Siegfried Handschuh}, \bibinfo{person}{Andr{\'e} Freitas}, \bibinfo{person}{Brian Davis}, \bibinfo{person}{Ross McDermott}, \bibinfo{person}{Manel Zarrouk}, {and} \bibinfo{person}{Alexandra Balahur}.} \bibinfo{year}{2018}\natexlab{}.
\newblock \showarticletitle{Www'18 open challenge: financial opinion mining and question answering}. In \bibinfo{booktitle}{\emph{Companion proceedings of the the web conference 2018}}. \bibinfo{pages}{1941--1942}.
\newblock


\bibitem[Merrick et~al\mbox{.}(2024)]%
        {merrick2024arctic}
\bibfield{author}{\bibinfo{person}{Luke Merrick}, \bibinfo{person}{Danmei Xu}, \bibinfo{person}{Gaurav Nuti}, {and} \bibinfo{person}{Daniel Campos}.} \bibinfo{year}{2024}\natexlab{}.
\newblock \showarticletitle{Arctic-Embed: Scalable, Efficient, and Accurate Text Embedding Models}.
\newblock \bibinfo{journal}{\emph{arXiv preprint arXiv:2405.05374}} (\bibinfo{year}{2024}).
\newblock


\bibitem[Mitra et~al\mbox{.}(2017)]%
        {mitra2017learning}
\bibfield{author}{\bibinfo{person}{Bhaskar Mitra}, \bibinfo{person}{Fernando Diaz}, {and} \bibinfo{person}{Nick Craswell}.} \bibinfo{year}{2017}\natexlab{}.
\newblock \showarticletitle{Learning to match using local and distributed representations of text for web search}. In \bibinfo{booktitle}{\emph{Proceedings of the 26th international conference on world wide web}}. \bibinfo{pages}{1291--1299}.
\newblock


\bibitem[Moreira et~al\mbox{.}(2024)]%
        {moreira2024nv}
\bibfield{author}{\bibinfo{person}{Gabriel de Souza~P Moreira}, \bibinfo{person}{Radek Osmulski}, \bibinfo{person}{Mengyao Xu}, \bibinfo{person}{Ronay Ak}, \bibinfo{person}{Benedikt Schifferer}, {and} \bibinfo{person}{Even Oldridge}.} \bibinfo{year}{2024}\natexlab{}.
\newblock \showarticletitle{NV-Retriever: Improving text embedding models with effective hard-negative mining}.
\newblock \bibinfo{journal}{\emph{arXiv preprint arXiv:2407.15831}} (\bibinfo{year}{2024}).
\newblock


\bibitem[Muennighoff et~al\mbox{.}(2022)]%
        {muennighoff2022mteb}
\bibfield{author}{\bibinfo{person}{Niklas Muennighoff}, \bibinfo{person}{Nouamane Tazi}, \bibinfo{person}{Lo{\"\i}c Magne}, {and} \bibinfo{person}{Nils Reimers}.} \bibinfo{year}{2022}\natexlab{}.
\newblock \showarticletitle{MTEB: Massive text embedding benchmark}.
\newblock \bibinfo{journal}{\emph{arXiv preprint arXiv:2210.07316}} (\bibinfo{year}{2022}).
\newblock


\bibitem[Nogueira and Cho(2019)]%
        {nogueira2019passage}
\bibfield{author}{\bibinfo{person}{Rodrigo Nogueira} {and} \bibinfo{person}{Kyunghyun Cho}.} \bibinfo{year}{2019}\natexlab{}.
\newblock \showarticletitle{Passage Re-ranking with BERT}.
\newblock \bibinfo{journal}{\emph{arXiv preprint arXiv:1901.04085}} (\bibinfo{year}{2019}).
\newblock


\bibitem[Nogueira et~al\mbox{.}(2019)]%
        {nogueira2019multi}
\bibfield{author}{\bibinfo{person}{Rodrigo Nogueira}, \bibinfo{person}{Wei Yang}, \bibinfo{person}{Kyunghyun Cho}, {and} \bibinfo{person}{Jimmy Lin}.} \bibinfo{year}{2019}\natexlab{}.
\newblock \showarticletitle{Multi-stage document ranking with BERT}.
\newblock \bibinfo{journal}{\emph{arXiv preprint arXiv:1910.14424}} (\bibinfo{year}{2019}).
\newblock


\bibitem[Oord et~al\mbox{.}(2018)]%
        {oord2018representation}
\bibfield{author}{\bibinfo{person}{Aaron van~den Oord}, \bibinfo{person}{Yazhe Li}, {and} \bibinfo{person}{Oriol Vinyals}.} \bibinfo{year}{2018}\natexlab{}.
\newblock \showarticletitle{Representation learning with contrastive predictive coding}.
\newblock \bibinfo{journal}{\emph{arXiv preprint arXiv:1807.03748}} (\bibinfo{year}{2018}).
\newblock


\bibitem[Qiao et~al\mbox{.}(2019)]%
        {qiao2019understanding}
\bibfield{author}{\bibinfo{person}{Yifan Qiao}, \bibinfo{person}{Chenyan Xiong}, \bibinfo{person}{Zhenghao Liu}, {and} \bibinfo{person}{Zhiyuan Liu}.} \bibinfo{year}{2019}\natexlab{}.
\newblock \showarticletitle{Understanding the Behaviors of BERT in Ranking}.
\newblock \bibinfo{journal}{\emph{arXiv preprint arXiv:1904.07531}} (\bibinfo{year}{2019}).
\newblock


\bibitem[Ram et~al\mbox{.}(2023)]%
        {ram2023context}
\bibfield{author}{\bibinfo{person}{Ori Ram}, \bibinfo{person}{Yoav Levine}, \bibinfo{person}{Itay Dalmedigos}, \bibinfo{person}{Dor Muhlgay}, \bibinfo{person}{Amnon Shashua}, \bibinfo{person}{Kevin Leyton-Brown}, {and} \bibinfo{person}{Yoav Shoham}.} \bibinfo{year}{2023}\natexlab{}.
\newblock \showarticletitle{In-context retrieval-augmented language models}.
\newblock \bibinfo{journal}{\emph{Transactions of the Association for Computational Linguistics}}  \bibinfo{volume}{11} (\bibinfo{year}{2023}), \bibinfo{pages}{1316--1331}.
\newblock


\bibitem[Reimers and Gurevych(2019)]%
        {reimers2019sentence}
\bibfield{author}{\bibinfo{person}{Nils Reimers} {and} \bibinfo{person}{Iryna Gurevych}.} \bibinfo{year}{2019}\natexlab{}.
\newblock \showarticletitle{Sentence-bert: Sentence embeddings using siamese bert-networks}.
\newblock \bibinfo{journal}{\emph{arXiv preprint arXiv:1908.10084}} (\bibinfo{year}{2019}).
\newblock


\bibitem[Thakur et~al\mbox{.}(2021)]%
        {thakur2021beir}
\bibfield{author}{\bibinfo{person}{Nandan Thakur}, \bibinfo{person}{Nils Reimers}, \bibinfo{person}{Andreas R{\"u}ckl{\'e}}, \bibinfo{person}{Abhishek Srivastava}, {and} \bibinfo{person}{Iryna Gurevych}.} \bibinfo{year}{2021}\natexlab{}.
\newblock \showarticletitle{Beir: A heterogenous benchmark for zero-shot evaluation of information retrieval models}.
\newblock \bibinfo{journal}{\emph{arXiv preprint arXiv:2104.08663}} (\bibinfo{year}{2021}).
\newblock


\bibitem[Vaswani et~al\mbox{.}(2017)]%
        {vaswani2017attention}
\bibfield{author}{\bibinfo{person}{Ashish Vaswani}, \bibinfo{person}{Noam Shazeer}, \bibinfo{person}{Niki Parmar}, \bibinfo{person}{Jakob Uszkoreit}, \bibinfo{person}{Llion Jones}, \bibinfo{person}{Aidan~N Gomez}, \bibinfo{person}{{\L}ukasz Kaiser}, {and} \bibinfo{person}{Illia Polosukhin}.} \bibinfo{year}{2017}\natexlab{}.
\newblock \showarticletitle{Attention is all you need}.
\newblock \bibinfo{journal}{\emph{Advances in neural information processing systems}}  \bibinfo{volume}{30} (\bibinfo{year}{2017}).
\newblock


\bibitem[Wang et~al\mbox{.}(2022a)]%
        {wang2022simlm}
\bibfield{author}{\bibinfo{person}{Liang Wang}, \bibinfo{person}{Nan Yang}, \bibinfo{person}{Xiaolong Huang}, \bibinfo{person}{Binxing Jiao}, \bibinfo{person}{Linjun Yang}, \bibinfo{person}{Daxin Jiang}, \bibinfo{person}{Rangan Majumder}, {and} \bibinfo{person}{Furu Wei}.} \bibinfo{year}{2022}\natexlab{a}.
\newblock \showarticletitle{Simlm: Pre-training with representation bottleneck for dense passage retrieval}.
\newblock \bibinfo{journal}{\emph{arXiv preprint arXiv:2207.02578}} (\bibinfo{year}{2022}).
\newblock


\bibitem[Wang et~al\mbox{.}(2022b)]%
        {wang2022text}
\bibfield{author}{\bibinfo{person}{Liang Wang}, \bibinfo{person}{Nan Yang}, \bibinfo{person}{Xiaolong Huang}, \bibinfo{person}{Binxing Jiao}, \bibinfo{person}{Linjun Yang}, \bibinfo{person}{Daxin Jiang}, \bibinfo{person}{Rangan Majumder}, {and} \bibinfo{person}{Furu Wei}.} \bibinfo{year}{2022}\natexlab{b}.
\newblock \showarticletitle{Text embeddings by weakly-supervised contrastive pre-training}.
\newblock \bibinfo{journal}{\emph{arXiv preprint arXiv:2212.03533}} (\bibinfo{year}{2022}).
\newblock


\bibitem[Wang et~al\mbox{.}(2023)]%
        {wang2023improving}
\bibfield{author}{\bibinfo{person}{Liang Wang}, \bibinfo{person}{Nan Yang}, \bibinfo{person}{Xiaolong Huang}, \bibinfo{person}{Linjun Yang}, \bibinfo{person}{Rangan Majumder}, {and} \bibinfo{person}{Furu Wei}.} \bibinfo{year}{2023}\natexlab{}.
\newblock \showarticletitle{Improving text embeddings with large language models}.
\newblock \bibinfo{journal}{\emph{arXiv preprint arXiv:2401.00368}} (\bibinfo{year}{2023}).
\newblock


\bibitem[Wang et~al\mbox{.}(2020a)]%
        {wang2020minilmv2}
\bibfield{author}{\bibinfo{person}{Wenhui Wang}, \bibinfo{person}{Hangbo Bao}, \bibinfo{person}{Shaohan Huang}, \bibinfo{person}{Li Dong}, {and} \bibinfo{person}{Furu Wei}.} \bibinfo{year}{2020}\natexlab{a}.
\newblock \showarticletitle{Minilmv2: Multi-head self-attention relation distillation for compressing pretrained transformers}.
\newblock \bibinfo{journal}{\emph{arXiv preprint arXiv:2012.15828}} (\bibinfo{year}{2020}).
\newblock


\bibitem[Wang et~al\mbox{.}(2020b)]%
        {wang2020minilm}
\bibfield{author}{\bibinfo{person}{Wenhui Wang}, \bibinfo{person}{Furu Wei}, \bibinfo{person}{Li Dong}, \bibinfo{person}{Hangbo Bao}, \bibinfo{person}{Nan Yang}, {and} \bibinfo{person}{Ming Zhou}.} \bibinfo{year}{2020}\natexlab{b}.
\newblock \showarticletitle{Minilm: Deep self-attention distillation for task-agnostic compression of pre-trained transformers}.
\newblock \bibinfo{journal}{\emph{Advances in Neural Information Processing Systems}}  \bibinfo{volume}{33} (\bibinfo{year}{2020}), \bibinfo{pages}{5776--5788}.
\newblock


\bibitem[Yang et~al\mbox{.}(1903)]%
        {yang1903simple}
\bibfield{author}{\bibinfo{person}{Wei Yang}, \bibinfo{person}{Haotian Zhang}, {and} \bibinfo{person}{Jimmy Lin}.} \bibinfo{year}{1903}\natexlab{}.
\newblock \showarticletitle{Simple applications of BERT for ad hoc document retrieval. CoRR abs/1903.10972 (2019)}.
\newblock \bibinfo{journal}{\emph{URL: http://arxiv. org/abs/1903.10972}} (\bibinfo{year}{1903}).
\newblock


\bibitem[Yang et~al\mbox{.}(2018)]%
        {yang2018hotpotqa}
\bibfield{author}{\bibinfo{person}{Zhilin Yang}, \bibinfo{person}{Peng Qi}, \bibinfo{person}{Saizheng Zhang}, \bibinfo{person}{Yoshua Bengio}, \bibinfo{person}{William~W Cohen}, \bibinfo{person}{Ruslan Salakhutdinov}, {and} \bibinfo{person}{Christopher~D Manning}.} \bibinfo{year}{2018}\natexlab{}.
\newblock \showarticletitle{HotpotQA: A dataset for diverse, explainable multi-hop question answering}.
\newblock \bibinfo{journal}{\emph{arXiv preprint arXiv:1809.09600}} (\bibinfo{year}{2018}).
\newblock


\bibitem[Zamani et~al\mbox{.}(2018)]%
        {zamani2018neural}
\bibfield{author}{\bibinfo{person}{Hamed Zamani}, \bibinfo{person}{Mostafa Dehghani}, \bibinfo{person}{W~Bruce Croft}, \bibinfo{person}{Erik Learned-Miller}, {and} \bibinfo{person}{Jaap Kamps}.} \bibinfo{year}{2018}\natexlab{}.
\newblock \showarticletitle{From neural re-ranking to neural ranking: Learning a sparse representation for inverted indexing}. In \bibinfo{booktitle}{\emph{Proceedings of the 27th ACM international conference on information and knowledge management}}. \bibinfo{pages}{497--506}.
\newblock


\end{thebibliography}

%\appendix

\end{document}